\documentstyle[aps,amsfonts,prd]{revtex}

\begin{document}

\title {One-loop graviton corrections to Maxwell's equations}

\author{Diego A.R. Dalvit $^{1}$ \thanks{dalvit@lanl.gov},
Francisco D. Mazzitelli $^{2}$ \thanks{fmazzi@df.uba.ar}, and
Carmen Molina-Par\'{\i}s $^{3}$ \thanks{molina@laeff.esa.es} }

\address{$^1$ T-6 Theoretical Division, MS B288, Los Alamos National
Laboratory, Los Alamos, NM 87545}

\address{$^2$
Departamento de F\'\i sica {\it J.J. Giambiagi},
Facultad de Ciencias Exactas y Naturales\\
Universidad de Buenos Aires - Ciudad Universitaria, Pabell\' on I,
1428 Buenos Aires, Argentina}

\address{$^3$
Centro de Astrobiolog\'{\i}a, CSIC--INTA, Carretera de Ajalvir Km. 4,
28850 Torrej\'on, Madrid, Spain}

\maketitle

\begin{abstract}

We compute the graviton induced corrections to Maxwell's equations in
the one-loop and weak field approximations.  The corrected equations
are analogous to the classical equations in anisotropic and
inhomogeneous media.  We analyze in particular the corrections to the
dispersion relations.  When the wavelength of the electromagnetic
field is much smaller than a typical length scale of the graviton
two-point function, the speed of light depends on the direction of
propagation and on the polarisation of the radiation.  In the opposite
case, the speed of light may also depend on the energy of the
electromagnetic radiation.  We study in detail wave propagation in two
special backgrounds, flat Robertson-Walker and static, spherically
symmetric spacetimes. In the case of a flat Robertson-Walker
gravitational background we find that the corrected electromagnetic
field equations correspond to an isotropic medium with a
time-dependent effective refractive index.  For a static, spherically
symmetric background the graviton fluctuations induce a vacuum
structure which causes birefringence in the propagation of light.

\medskip

{PACS number(s): 04.60.-m, 11.15.Kc.}

\end{abstract}


\section{INTRODUCTION}
\label{intro}

It is well known that when the QED vacuum is modified by external
conditions such as background fields, finite temperature, or boundary
conditions, the propagation of the photons can be affected in a
non-trivial way.  The vacuum behaves as a dispersive medium in which
the propagation of light generally depends on the direction of
propagation and on the polarisation of the radiation. Physically, the
effect can be understood as follows: the photon exists for part of the
time as a virtual $e^-e^+$ pair, on which the external conditions do
act and modify the propagation.  In previous works the problem of
photon propagation in modified QED vacua has been analyzed for
external electromagnetic (EM) fields~\cite{adler,dittrich}, boundary
conditions~\cite{scharn1,barton}, external gravitational
fields~\cite{dh,danshore}, and finite
temperature~\cite{tarrach}. Further references and details can be
found in Ref.~\cite{scharn2}. There is also an experiment under
construction to detect birefringence of the QED vacuum in the presence
of a strong magnetic field~\cite{roberto}.

The phenomenon is of course quite general, and not restricted to
$e^-e^+ $ pairs. The interaction of the photon with any other field
will produce similar effects: the photon will not follow, in general,
a geodesic of spacetime~\footnote {It is even possible to have
superluminal propagation. However, as extensively discussed in the
literature~\cite{dh,scharn2,shore}, this does not imply causality
violations.}.  In this paper we will analyze the effect of the
coupling between a classical electromagnetic field and a quantum
gravitational field on the propagation of EM radiation waves.  We will
show that, indeed, the graviton loop leads to effects that are similar
to those already studied and calculated for QED vacua.

Even though General Relativity is a non-renormalizable theory, the
one-loop corrections are meaningful when the quantized gravitational
field theory is looked upon as an effective field
theory~\cite{donoghue}. It is possible to compute, for instance, the
leading (long distance) quantum corrections to the Newtonian
potential~\cite{newtpot,dddm}.  Our calculation provides another
example of a quantum gravity effect that can be estimated using
General Relativity as a low energy effective field theory for quantum
gravity. Moreover, it could also be of some interest from a
phenomenological point of view. Indeed, Amelino-Camelia {\it et
al}~\cite{nature} pointed out that many quantum gravity scenarios
predict a frequency-dependent velocity of light that could be
observable for (cosmological) gamma-ray bursts. Gambini and
Pullin~\cite{gambinipullin} studied the propagation of light in
canonical quantum gravity and found that the modified Maxwell's
equations imply a frequency and helicity dependent velocity of
propagation.  We will see that, in principle, similar results can be
found by taking into account the interaction between gravitons and EM
radiation in the low energy theory.

The paper is organized as follows. In Section II we obtain the quantum
corrections to the classical Maxwell's equations induced by the
graviton loop. We write the corrections in terms of the coincidence
limit of the graviton two-point function. In Section III we present a
qualitative analysis of the quantum corrections to the dispersion
relations.  We analyze the cases $\lambda >> L_c$ and $\lambda <<
L_c$, where $\lambda$ is the wavelength of the classical
electromagnetic radiation and $L_c$ is a typical scale of variation of
the graviton two-point function. We show that in general the velocity
of light depends on the direction of propagation and on the
polarisation, {\it i.e.}, we find gravitational birefringence. In
Section IV we discuss quantitatively two examples: a flat
Robertson-Walker (RW) background metric and a static metric with
spherical symmetry. Section V contains our conclusions and final
remarks.


\section{EFFECTIVE EQUATIONS OF MOTION}
\label{equations}

Consider pure gravity described by the Einstein-Hilbert (classical)
action~\footnote{Our metric has signature $(-+++)$ and the Riemann and
Ricci tensors, and the scalar curvature are defined as
${R}^{\mu}_{~\nu\alpha\beta} = \partial_{\alpha}
\Gamma^{\mu}_{\nu\beta} - \ldots$, ${R}_{\alpha\beta} =
{R}^{\mu}_{~\alpha\mu\beta}$, and ${R}= {g}^{\alpha\beta}
{R}_{\alpha\beta}$, respectively. We use units such that $\hbar=c=1$.}

\begin{eqnarray}
S_{\rm G}=\frac{2}{\kappa^2}
\int {\rm d}^4x \sqrt{-g} R
\; ,
\label{Gaction}
\end{eqnarray}
where $\kappa^2=32 \pi G$ and $R$ is the Ricci scalar.  The classical
action for the EM field is given by

\begin{eqnarray}
S_{\rm EM}=-\frac{1}{4} \int {\rm d}^4x \sqrt{-g} F_{\mu\nu} F^{\mu\nu}
\; ,
\label{EMaction}
\end{eqnarray}
where $F_{\mu\nu}=\nabla_{\mu} A_{\nu} - \nabla_{\nu} A_{\mu}$ is the
field strength tensor and $A_\mu$ the gauge potential.  The classical
energy-momentum tensor associated to the EM field is given by

\begin{eqnarray}
T^{\sigma \tau}_{\rm EM} = F^{\sigma}_{\; \;\mu} F^{\mu \tau} -
\frac{1}{4} g^{\sigma \tau} F_{\mu\nu} F^{\mu\nu}
\; .
\end{eqnarray}
The classical action of the EM field depends on the (classical)
gravitational background. It is then natural to ask ourselves what are
the effects on Maxwell's equations due to a change in the
gravitational background, and in this paper particularly, what are the
implications of the one-loop graviton fluctuations.

We define the classical action of the combined system (gravitational
field plus classical EM radiation)

\begin{eqnarray}
S_{\rm clas}&=& S_{\rm EM} + S_{G}
\; .
\end{eqnarray}

The effect of quantum metric fluctuations can be analyzed by means of
the background field method, expanding the total action $S_{\rm clas}$
around a background metric as $g_{\mu\nu} \rightarrow g_{\mu\nu} +
\kappa h_{\mu\nu}$, and integrating over the graviton field
($h_{\mu\nu}$) degrees of freedom to get an effective action for the
background fields $g_{\mu\nu}$ and $A_{\mu}$.  In order to fix the
gauge one must choose a gauge-breaking term $\chi^{\mu}[g,h]$, with
its corresponding gauge-breaking action $S_{\rm gauge}[g,h]= -(1/2)
\int {\rm d}^4x \sqrt{-g} \chi^{\mu} g_{\mu\nu} \chi^{\nu}$, and its
corresponding ghost action $S_{\rm ghost}$~\cite{dewitt}. The complete
effective action $S_{\rm eff}$ is obtained by integrating the full
action $S\equiv S_{\rm clas} + S_{\rm gauge} + S_{\rm ghost}$ over the
graviton and ghost fields. We evaluate this effective action in the
one-loop approximation, for which we expand $S$ up to second order in
gravitons. The second order term can be shown to have the
form~\cite{dd}

\begin{equation}
S^{(2)}=\int {\rm d}^4x \sqrt{-g} \; h_{\mu\nu} ( O^{\mu\nu\sigma\tau} +
P^{\mu\nu\sigma\tau} ) h_{\sigma\tau}
\; ,
\end{equation}
where $\hat{O}\equiv O^{\mu\nu\sigma\tau}$ is a second order
differential operator that depends on the background metric and is
independent of the EM field~\cite{dd} (we will not need its exact form
in what follows), and the tensor $\hat{P} \equiv P^{\mu\nu\sigma\tau}$
arises from the expansion of $S_{\rm EM}$ to second order in
gravitons, and reads

\begin{eqnarray}
P^{\mu\nu\sigma\tau} &=&
-\frac{\kappa^2}{8} \left[
F_{\alpha\beta} F^{\alpha \beta}
\left( \frac{1}{4} g^{\mu\nu} g^{\sigma\tau} -
\frac{1}{2} g^{\mu\sigma} g^{\nu\tau} \right) +
\frac{1}{2} F_{\alpha\beta} g^{\mu\nu}
\left( -g^{\beta\sigma} F^{\alpha\tau} + g^{\alpha\sigma} F^{\beta\tau}
\right)
 \right. \nonumber \\
&+& \left. \frac{1}{2} F_{\alpha\beta} g^{\sigma \tau}
\left( -g^{\beta\mu} F^{\alpha\nu} + g^{\alpha\mu} F^{\beta\nu} \right)
+
2 F_{\alpha\beta}
\left( F^{\alpha\tau} g^{\beta\mu} g^{\nu\sigma} +
       F^{\nu\beta} g^{\alpha\sigma} g^{\mu\tau} +
       F^{\nu\tau} g^{\alpha\mu} g^{\beta\sigma} \right)
\right]
\; .
\end{eqnarray}
There is also a second order term in the ghost fields that for
gauge-breaking terms linear in the metric fluctuations decouples from
the gravitons and is only coupled to the background metric. This means
that the one-loop effective action for the combined system
reads~\cite{dd}

\begin{equation}
S_{\rm eff}[g_{\mu\nu},A_{\mu}] = S_{\rm clas} + \frac{i}{2}
{\rm{Tr}} \ln( \hat{O} + \hat{P}) - i {\rm{Tr}} \ln \hat{G}_{\rm ghost}
\; ,
\end{equation}
where $\hat{G}_{\rm ghost}$ is the second order differential operator
that comes from integrating over the ghost fields.

It is extremely complicated, in general, to calculate the one-loop
effective action. In this paper we will use the weak field
approximation, assuming that the EM field is a test field that does
not affect the background metric $g_{\mu \nu}$. In this approximation
the effective action takes the form

\begin{equation}
S_{\rm eff} = S_{\rm clas} + \frac{i}{2} {\rm{Tr}} \ln \hat{O} - i
{\rm{Tr}}
\ln \hat{G}_{\rm ghost}  +
\int {\rm d}^4x \sqrt{-g} F_{\mu\nu}(x) F_{\sigma\tau}(x)
M^{\mu\nu\sigma\tau}(x,x)
\; ,
\label{efectiva}
\end{equation}
where

\begin{eqnarray}
M^{\mu\nu\sigma\tau}(x,x) &=& -\frac{\kappa^2}{16}
\langle h_{\alpha\beta}(x) h_{\zeta\eta}(x)
+ h_{\zeta\eta}(x) h_{\alpha\beta}(x)
\rangle
\left[
\left(
\frac{1}{8} g^{\alpha\beta} g^{\zeta \eta}
- \frac{1}{4} g^{\alpha \zeta} g^{\beta\eta}
\right)
(g^{\sigma\mu} g^{\tau\nu} -
g^{\sigma\nu} g^{\tau\mu})  \right. \nonumber \\
&+& \frac{1}{2} g^{\alpha\beta}
(g^{\mu \zeta} g^{\sigma\nu} g^{\tau\eta} -
 g^{\mu\sigma} g^{\nu \zeta} g^{\tau\eta} -
 g^{\sigma\eta} g^{\mu \zeta} g^{\tau\nu} +
 g^{\sigma\eta}  g^{\mu\tau} g^{\nu \zeta})  \nonumber \\
&+& g^{\alpha \zeta}
(g^{\mu\sigma} g^{\tau\eta} g^{\nu\beta} +
 g^{\sigma\beta} g^{\tau\nu} g^{\mu\eta} -
 g^{\nu\sigma} g^{\tau\eta} g^{\mu\beta} -
 g^{\sigma\beta}  g^{\tau\mu} g^{\nu\eta})  \nonumber \\
&+& \left. g^{\sigma\beta} g^{\tau\eta} g^{\mu\alpha} g^{\nu \zeta} -
g^{\sigma\beta} g^{\tau\eta} g^{\nu\alpha} g^{\mu \zeta}
\right]
\; .
\label{eme}
\end{eqnarray}
Note that the tensor $M^{\mu\nu\sigma\tau}$ has the following symmetry
properties, which are similar to the symmetries of the Riemann tensor:
$M^{\mu\nu\sigma\tau} = - M^{\nu\mu\sigma\tau}$, $M^{\mu\nu\sigma\tau}
= - M^{\mu\nu\tau\sigma}$, and $M^{\mu\nu\sigma\tau} =
M^{\sigma\tau\mu\nu}$.  In view of these properties, the only
non-vanishing components of this tensor are $M^{0i0j}$, $M^{0ijk}$,
and $M^{ijkl}$, where $i,j,k,l$ are spatial indices.  This tensor has
21 independent components and depends on the two-point function of
gravitons, evaluated in an arbitrary quantum state $\vert \Psi
\rangle$.  The two-point function of gravitons can be written as
follows

\begin{equation}
G^{\mu\nu\sigma\tau} (x,x') \equiv
\langle \Psi | h^{\mu\nu}(x) h^{\sigma\tau}(x') + h^{\sigma\tau}(x')
h^{\mu\nu}(x) | \Psi \rangle
\; ,
\label{2point}
\end{equation}
taken at coincidence $(x=x')$. In the following we will assume that
the graviton state $\vert \Psi \rangle$ preserves the symmetries of
the background metric, so that the tensor $M^{\mu\nu\sigma\tau}$ will
share those same symmetries.

The (one-loop) gravitationally modified equations of motion for the EM
field, $\delta S_{\rm eff}/\delta A_{\nu}=0,$ are given by

\begin{eqnarray}
\nabla_\mu G^{\mu \nu}=0 \; ,
  \;  &  & \;  \;
G^{\mu \nu } =
 F^{\mu\nu} - 4  M^{\mu\nu}_{\;\;\;\sigma\tau}
F^{\sigma\tau}
\; .
\label{maxwell}
\end{eqnarray}
These equations are analogous to the classical Maxwell's equations in
the presence of a linear, anisotropic, and non-homogeneous media. To
see this explicitly, we recall that in a local Lorentz frame
$F^{0m}=E^m$ and $F^{mn}=\epsilon^{mn}_{\;\;\;\;k} B^k$, so that we
can introduce the vectors $G^{0m}=D^m$ and
$G^{mn}=\epsilon^{mn}_{\;\;\;\;k} H^k$, namely

\begin{mathletters}
\begin{eqnarray}
D^j &\equiv & E^j - 8 M^{0j}_{~~0m}E^m -
4 M^{0j}_{~~mn}\epsilon^{mn}_{~~k}   B^k
\; ,
\\
H^j &\equiv& B^j- 2 \epsilon^{jmn} M_{mnik}
\epsilon^{ik}_{~~q} B^q
- 4 \epsilon^{jki} M_{ki0m} E^m
\; .
\end{eqnarray}
\end{mathletters}
The quantum corrected equations in a local Lorentz frame read

\begin{mathletters}
\begin{eqnarray}
{\bf  \nabla}
 \cdot {\bf D} &=&0
\; ,
\\
{\bf  \nabla} \cdot {\bf B} &=&0
\; ,
\\
{\bf  \nabla} \times {\bf H} &=&
\partial_t {\bf D}
\; ,
\\
{\bf  \nabla} \times {\bf E} &=&
-\partial_t {\bf B}
\; ,
\label{maxeq}
\end{eqnarray}
\end{mathletters}
where we have also included the equations that follow from the Bianchi
identity $\nabla_\mu F_{\nu \sigma} + \nabla_\nu F_{\sigma \mu} +
\nabla_\sigma F_{\mu \nu} =0$.  The constitutive relations ${\bf
D}[{\bf E},{\bf B}]$ and ${\bf H}[{\bf E},{\bf B}]$ are unusual when
$M_{0ijk}\neq0$, since in this case {\it both} the electric and
magnetic fields appear in the definition of ${\bf D}$ and ${\bf
H}$. However, when $M_{0ijk}=0$ the relations are exactly equivalent
to those of a linear medium characterized by spacetime dependent
electric permitivity and magnetic permeability tensors defined as

\begin{mathletters}
\begin{eqnarray}
D^i = \epsilon^i_{~j} E^j \; ,  & \;\;  \;\; &
\epsilon^i_{~j} \equiv \delta^i_{~j} - 8 M^{i0}_{~~0j} \; ,
\\
B^i = \mu^i_{~j} H^j \; , & \;\;  \;\; &
\mu^i_{~j}  \equiv \delta^i_{~j} + 2 \epsilon^i_{~mn}
M^{mnab} \epsilon_{abj}
\; .
\label{pertensors}
\end{eqnarray}
\end{mathletters}

We emphasize two relevant technical points. On the one hand, the
graviton two-point function will depend, in general, on the choice of
the gauge-breaking term for the graviton fluctuations.  However, the
background metric will also depend on the gauge-breaking term through
the semi-classical Einstein equation.  Both dependencies should cancel
out in the dynamics of any test field (see Refs.~\cite{dddm}
and~\cite{sidm}).  We will not consider this problem in what follows,
{\it i.e.,} we will assume that the background metric already contains
the back reaction of gravitons, computed with a given gauge-breaking
term.  On the other hand, the two-point function will diverge at the
coincidence limit. Adequate counterterms are needed to absorb the
divergences. In the spirit of effective field theories, these
counterterms will not contribute in the long distance/low energy
limit, which will be dominated by the non-local, non-analytic part of
the two-point function~\cite{donoghue,dddm}.


\section{QUALITATIVE ANALYSIS}
\label{qualitative}

Given the quantum corrected equations, we can distinguish two
different physical regimes depending on the relative size of the
wavelength $\lambda$ of an EM radiation field described by
$F^{\mu\nu}$ and the typical scale of variation of
$M^{\mu\nu\sigma\tau}$, $L_c$.

When $\lambda \ll L_c$, we can take the tensor $M^{\mu\nu\sigma\tau}$
outside the covariant derivative in Eq. (\ref{maxwell}), as it does
not change significantly over the scale $\lambda$. The equation of
motion can then be written in the form

\begin{equation}
\nabla_{\mu} F^{\mu\nu} - 4 M^{\mu\nu}_{~~\sigma\tau}
\nabla_{\mu} F^{\sigma\tau} = 0
\; .
\label{maxapprox}
\end{equation}
We first introduce the following variables~\cite{dh}

\begin{eqnarray}
F_{\mu \nu}= f_{\mu \nu}e^{i \psi} \; ,
\end{eqnarray}
with $f_{\mu \nu}$ the amplitude and $\psi$ the phase, such that
$k_\mu =\nabla_\mu \psi$ is the momentum of the EM wave. We assume
that the amplitude $f_{\mu \nu}$ is the slow varying variable and
$\psi$ is the fast varying variable, so that from now on we discard
any gradients and/or time variations of the amplitude $f_{\mu \nu}$.

We start from equation (\ref{maxapprox}) and make use of the new
variables to write

\begin{eqnarray}
k_\rho f^{\rho\nu}
- 4 k_{\rho} M^{\rho \nu \sigma \tau} f_{\sigma \tau}&=&0
\; .
\label{max1}
\end{eqnarray}
The remaining Maxwell's equation implies the following

\begin{equation}
k_\mu f_{\nu \sigma} +
k_\nu f_{\sigma \mu} +
k_\sigma f_{\mu \nu}
=0
\; .
\label{max2}
\end{equation}
We now multiply equation (\ref{max1}) by $k^\mu$ and make use of
equation (\ref{max2}) to obtain

\begin{eqnarray}
k^2 f^{\mu \nu}
+ 8 k^{[\mu} M^{\nu] \sigma \tau \rho} k_\sigma f_{\tau \rho}&=&0
\; .
\label{dhnew}
\end{eqnarray}
This equation is similar to the one discussed by Drummond and Hathrell
in Ref.~\cite{dh}. (See also Ref.~\cite{danshore}.) In those
references the corrections are due to fermion loops and are
proportional to the Riemann tensor $R^{\mu\nu\rho\sigma}$ (this is
true in the case of empty spacetimes, so that $R^{\mu \nu}$ and $R$
vanish~\cite{dh}), instead of $M^{\mu\nu\rho\sigma}$.

In the absence of quantum corrections one obtains the usual dispersion
relation $k^2=0$. The graviton loop induces modifications to this
relation, {\it i.e.}, light rays do not follow null spacetime
geodesics. We will analyse in detail several examples in the following
sections. Here we discuss qualitatively some general properties of the
modified dispersion relations.

It is easy to see that the tensor $M^{\mu\nu\rho\sigma}$ is
dimensionless and proportional to the square of the Planck length $L_P
\sim \kappa$. Therefore we expect $M^{\mu\nu\rho\sigma}=(L_P /L_c)^2
C^{\mu\nu\rho\sigma}$, with $C^{\mu\nu\rho\sigma}$ a dimensionless,
slowly varying tensor. Inserting this in Eq. (\ref{dhnew}), we see
that the modified dispersion relation will have the general form

\begin{equation}
k^2 + c_{\mu\nu}k^{\mu}k^{\nu} = 0
\; ,
\end{equation}
where $c_{\mu\nu}$ is a slowly varying tensor of order $O(L_P^2 /
L_c^2)$ that depends on the direction of propagation and the
polarisation of the EM radiation. Therefore, we expect the
modifications in the speed of light to be proportional to
${L_P^2/L_c^2}$ and gravitational birefringence of the same order of
magnitude.

We consider now the opposite case $\lambda \gg L_c$. For simplicity,
and in order to be able to compare with previous works, we assume that the
background metric is flat and that the quantum state for gravitons is
such that the two-point function has a random variation on micro-scales
(much smaller than any other scale of the system, but still larger
than $L_P$). In other words, we assume that the spacetime looks
classical at scales larger than $L_c$ and has a complicated random
structure at scales smaller than $L_c$. This kind of states has been
considered before in the context of loop quantum
gravity~\cite{gambinipullin,weave}.

In this situation, the quantum corrected equation (\ref{maxwell})
reads

\begin{equation}
 \partial_{\mu} F^{\mu\nu} - 4 M^{\mu\nu}_{~~\sigma\tau}
(\partial_{\mu} F^{\sigma\tau})
 - 4 (\partial_{\mu} M^{\mu\nu}_{~~\sigma\tau})
F^{\sigma\tau}
 = 0
\; .
\label{max3}
\end{equation}
It is, of course, not possible to neglect the derivatives of the
tensor $M_{\mu\nu\rho\sigma}$. The equation describes the propagation
of a classical electromagnetic wave in a random media.

As the wavelength is much larger than $L_c$, in order to obtain a
modified dispersion relation we average the field equation over a
spacetime domain of size $L^4$, with $L_c<<L<<\lambda$.  To compute
the average of the products $(\partial M) F$ and $M(\partial F)$ we
expand $F$ around a point $x_0$ in the domain. Schematically $F(x)
\approx F(x_0) + \partial F(x_0) (x - x_0)$. Denoting by $\langle
... \rangle$ the average over the domain and using that for a random
structure $\langle M \rangle = 0$ and $\langle \partial M \rangle =
0$, we obtain $ \langle F \partial M \rangle \approx \partial F(x_0)
\langle \partial M (x-x_0) \rangle$ and $\langle M \partial F \rangle
\approx \partial^2 F(x_0) \langle M (x-x_0) \rangle$. Therefore, in
this approximation, the average of Eq. (\ref{max3}) will contain
higher derivatives of the EM fields (as long as $\langle M (x-x_0)
\rangle\neq 0$).  As a consequence, on dimensional grounds we expect a
modified dispersion relation of the form

\begin{equation}
k^2 + c_{\mu\nu}k^{\mu}k^{\nu} + c_{\mu\nu\rho} k^{\mu}
k^{\nu}k^{\rho}= 0
\; ,
\end{equation}
where $c_{\mu\nu}= O(L_P^2 / L^2)$ and
$c_{\mu\nu\rho}= O(L_P^2 / L)$.

The quadratic correction modifies the speed of light as in the
previous case. The cubic term is qualitatively different, since it
produces a variation of the speed of light that increases linearly
with the energy of the EM radiation.  As already mentioned, this kind
of correction may be relevant from a phenomenological point of view,
because it induces a non-trivial structure in the arrival time of
light rays coming from gamma-ray bursts~\cite{nature,gambinipullin}.


\section{EXAMPLES}
\label{examples}

In the following sections we will concentrate on two particular
classes of gravitational backgrounds, namely flat RW metrics and
static spherically symmetric backgrounds.

\subsection{FLAT ROBERTSON-WALKER BACKGROUND}
\label{RW}

We first consider the case of a flat RW background, whose metric in
conformal coordinates reads~\footnote{For simplicity we have
considered the case of a flat RW background. Our results can be easily
generalized to the closed and open RW spacetimes.}

\begin{equation}
{\rm d} s^2= a^2(\eta) \left( -{\rm d} \eta^2 +
 {\rm d} {\bf{x}}^2 \right)
\; .
\end{equation}
Under the assumption that the graviton quantum vacuum preserves the
symmetries of the background metric, we can conclude that the tensor
$M^{\mu\nu\sigma\tau}$ has the same symmetries. For RW spacetimes the
metric is invariant under spatial reflections (due to its homogeneity
and isotropy), so that $M^{0ijk}=0$.  For the remaining two
non-vanishing sets of components of the tensor, $M^{0i0j}$ and
$M^{ijkl}$, we use the invariance of the metric under spatial
rotations. This implies that they can be written in the form

\begin{mathletters}
\begin{eqnarray}
M^{0i0j} &=& f_1(\eta) g^{ij} \; ,
\label{eme1}
\\
M^{ijkl}
&=& f_2(\eta) (g^{ik} g^{jl} - g^{il} g^{jk}) \; ,
\label{eme2}
\end{eqnarray}
\end{mathletters}
where $f_1(\eta)$ and $f_2(\eta)$ are functions of time. Note that the
non-vanishing components of the tensor have the same form as the
components of the Riemann tensor in RW spacetimes, apart from the
global factors $f_1$ and $f_2$. To determine these two functions, we
recall that the tensor $M_{\mu \nu \sigma \tau}$ is proportional to
the two-point function of gravitons, which in RW backgrounds can be
expressed in terms of the two-point function of a massless minimally
coupled scalar field~\cite{allen,fp}. Therefore the functions
$f_1(\eta)$ and $f_2(\eta)$ must be proportional to $\langle
\phi^2(\eta) \rangle$. In order to calculate the constants of
proportionality exactly, we need to go beyond symmetry arguments and
we must face the exact evaluation of the graviton two-point function
$G^{\mu\nu\rho\sigma}(x,x')$ of Eq. (\ref{2point}) in a RW
gravitational background. The result can be found in the literature
(see for example Ref.~\cite{allen}). We only need to quote the final
result of that reference. In the traceless ($h^{\mu}_{\; \mu}=0$),
transverse ($\nabla_{\mu} h^{\mu\nu}=0$), and synchronous gauge (hence
$h^{0\mu}=0$), and assuming the graviton vacuum state to be
homogeneous and isotropic, the coincidence limit of the two-point
function has only spatial non-vanishing components, and
reads~\footnote{The choice of vacuum corresponds to that used in
Ref.~\cite{allen}. This graviton vacuum is homogeneous, isotropic, and
the same for the two helicity states of the gravitons
$(+2,-2)$.}~\cite{allen}

\begin{equation}
G^{ijkl}(x,x)=2 \sum_{\bf k} ( m^i m^j m^{k*}
m^{l*} + m^{i*} m^{j*} m^{k} m^{l} ) |F(x,{\bf k})|^2
\; .
\label{coin}
\end{equation}
The sum over ${\bf k}$ denotes a sum over a three dimensional set of
spatial wave vectors. The complex (spacelike) vector $m^{i}({\bf k})$
is defined as $m^{i}({\bf k})= (1/{\sqrt{2}}) [e^{i}_1({\bf k}) + i
e^{i}_2({\bf k})]$. The vectors ${\bf e}_1({\bf k})$ and ${\bf
e}_2({\bf k})$ are spacelike vectors, such that the set $\{ {\bf
e}_1({\bf k}), {\bf e}_2({\bf k}), {\bf {\hat k}} \}$ forms an
orthonormal basis in the three dimensional spacelike hypersections
(here ${\bf {\hat k}}= {\bf k}/ k$ and $k=({\bf k} \cdot {\bf
k})^{1/2}$).  The mode function $F$ is given by $F(x,{\bf k})
=F(\eta,{\bf x},{\bf k}) \equiv f(\eta,{\bf k}) e^{i {\bf k}\cdot {\bf
x}} / \sqrt{32 \pi^3 k V}$, where $V$ is a constant comoving
volume~\footnote{The normalization of the mode function $F$ differs
from that of Allen~\cite{allen} in a $\kappa^{-1}$ factor. The reason
is that we have defined the graviton fluctuations via $g_{\mu\nu}
\rightarrow g_{\mu\nu} + \kappa h_{\mu\nu}$ and our graviton two-point
function, (see Eq. (\ref{2point})), is given in terms of this
$h^{\mu\nu}$, whereas Allen has $g_{\mu\nu} \rightarrow g_{\mu\nu} +
{\tilde h}_{\mu\nu} $, and defines the graviton two-point function in
terms of ${\tilde h}_{\mu\nu}$.}.  The mode functions $F(x,{\bf k})$
and $f({\bf k},\eta)$ are a solution to the equations

\begin{mathletters}
\begin{eqnarray}
&&\Box F(x,{\bf k})=0
\; ,
\\
&&\ddot f(\eta,{\bf k}) + \frac{2 \dot a}{a} \dot  f(\eta,{\bf k})
+ k^2 f(\eta,{\bf k})=0
\; ,
\end{eqnarray}
\end{mathletters}
respectively, {\it i.e.}, correspond to the dynamical equation of a
massless minimally coupled scalar field in a RW background. Here the
dot denotes derivation with respect to the conformal time variable
$\eta$.

The two-point function, given in Eq. (\ref{coin}), can be simplified
by making use of the identity $e^{i}_1 e^{j}_1 + e^{i}_2 e^{j}_2 +
k^{-2} k^{i} k^{j} = g^{ij}$. We can then perform the sum over momenta
as $|F(x,{\bf k})|^2$ depends only on the modulus of ${\bf k}$. The
final result reads

\begin{equation}
G^{ijkl}(x,x) = \frac{4}{15} \left(
\frac{3}{2} g^{ik} g^{jl} + \frac{3}{2}
g^{il} g^{jk} - g^{ij} g^{kl}
\right) \langle \phi^2(\eta) \rangle
\; ,
\label{cofin}
\end{equation}
where $\langle \phi^2(\eta) \rangle \equiv\sum_{{\bf k}} |F(x,{\bf
k})|^2$ is the coincidence limit of the two-point function of a
massless minimally coupled scalar field in a RW background. We can now
insert this formula in Eq. (\ref{eme}) in order to read off the
expressions for $f_1(\eta)$ and $f_2(\eta)$ in Eqs. (\ref{eme1}) and
(\ref{eme2}). This procedure yields

\begin{eqnarray}
f_1(\eta) &=&  \frac{\kappa^2}{48 a^2(\eta) }
\langle \phi^2(\eta) \rangle
\; ,
\nonumber \\
f_2(\eta) &=& -\frac{ \kappa^2}{16} \langle \phi^2(\eta)
\rangle
\; .
\end{eqnarray}
The quantum correction to the classical EM action due to the coupling
with the graviton degrees of freedom can now be obtained from
Eq. (\ref{efectiva}) by making use of the above results. We get

\begin{equation}
\langle S^{(2)}_{\rm EM}\rangle =
\int {\rm d}^4x \sqrt{-g} \;
 [- 4 a^2(\eta) f_1(\eta)  F_{0i} F^{0i} + 2 f_2(\eta) F_{ij} F^{ij}] 
\; .
\end{equation}
Note that the one-loop effective action for the electromagnetic field
is $S_{\rm EM} + \langle S^{(2)}_{\rm EM}\rangle$, which is a
divergent quantity and has to be suitably renormalized. This is
accomplished by the renormalization of $\langle\phi^2(\eta)\rangle$,
for example, by means of adiabatic regularization~\cite{bunch}.

Having the EM effective action and assuming that one first solves the
pure gravity part in order to get the corrected background metric
after graviton back reaction (that is, we assume one solves the pure
gravitational part and gets the new scale factor $a(\eta)$), one can
get the corrected Maxwell's equations from the variation of the
effective action, namely, $\delta S_{\rm eff}/\delta A_{\mu}=0$. We
will assume that $f_1(\eta)$ and $f_2(\eta)$ have a time variation
much slower than that associated with the EM field, so that we can
approximate the equations of motion as in Eq. (\ref{maxapprox}). The
source-free equations are the usual ones

\begin{mathletters}
\begin{eqnarray}
\frac{1}{a(\eta)} {\bf \nabla}\cdot {\bf B} &=& 0
\; ,
\\
\frac{1}{a(\eta)} {\dot{\bf B}} +
\frac{2 {\dot a}(\eta)}{a(\eta)} {\bf B} &=&
- \frac{1}{a(\eta)} {\bf \nabla} \times {\bf E}
\; ,
\end{eqnarray}
\end{mathletters}
where the dot denotes $\partial / \partial \eta$, and ${\bf \nabla}$
is vector notation for $\partial/\partial {\bf x}$.

The other two equations read
\begin{mathletters}
\begin{eqnarray}
\frac{1}{a(\eta)}  \epsilon_{\rm eff}(\eta) {\bf \nabla} \cdot {\bf E} &=&
0
\; ,
\\
\frac{1}{a(\eta)}  \epsilon_{\rm eff}(\eta) {\dot{\bf E}} +
\frac{2 {\dot a}(\eta)}{a(\eta)} \epsilon_{\rm eff}(\eta) {\bf E} &=&
\frac{1}{a(\eta)} \mu^{-1}_{\rm eff}(\eta)
{\bf \nabla} \times {\bf B}
\; .
\end{eqnarray}
\end{mathletters}
In the first term on the left hand side of the last equation we have
discarded a contribution proportional to the time derivative of the
effective electric permeability $\epsilon_{\rm eff}(\eta)$ since, as
already discussed in Eq. (\ref{maxapprox}), we are assuming that it
does not change significantly over the wavelength of the EM field.
The effective electric permeability and magnetic permitivity tensors
in RW backgrounds are proportional to the identity ($3 \times 3$)
matrix, namely, $\epsilon^i_{~j} = \epsilon_{\rm eff}(\eta)
\delta^i_{~j}$ and $\mu^i_{~j} = \mu_{\rm eff}(\eta) \delta^i_{~j}$,
with

\begin{mathletters}
\begin{eqnarray}
\epsilon_{\rm eff}(\eta) &\equiv& 1 + 8 a^2(\eta) f_1(\eta)
\; ,
 \\
\mu_{\rm eff}(\eta) &\equiv& 1+ 8 f_2(\eta) \; .
\end{eqnarray}
\end{mathletters}

Hence the presence of gravitons introduces a time dependent effective
refraction index $n_{\rm eff}(\eta) =\sqrt{\epsilon_{\rm eff}(\eta)
\mu_{\rm eff}(\eta) }$ for a travelling EM wave, and therefore implies
a time dependent speed of light in the medium~\footnote{This velocity
corresponds to the phase velocity $v_{\rm phase}=c/n_{\rm eff} $.}

\begin{equation}
v_{\rm eff}(\eta)  = 1 - 4 [a^2(\eta) f_1(\eta) + f_2(\eta)] =
1 + \frac{\kappa^2}{6} \langle \phi^2(\eta) \rangle
\; .
\label{veff}
\end{equation}
This effective speed of light is the same for all directions of
propagation and for all polarisations of the EM radiation field, in
agreement with the isotropy and homogeneity of RW spacetimes. In
Appendix \ref{rw-appendix} we give an alternative derivation of this
result based on a direct study of the dispersion relation for light in
the graviton modified medium.

The renormalized two-point function $\langle \phi^2(\eta) \rangle$
does not have a definite sign (see, for example, Refs.~\cite{dd,BD}).
The effective speed of light in the graviton vacuum can be greater or
smaller than that in free space, depending on the particular form of
the scaling parameter $a(\eta)$.  In any case, the correction is
extremely small, typically proportional to the ratio of the spacetime
scalar curvature and Planck's curvature $R/R_{\rm P}$.  As we go back
in time towards the Big Bang singularity, the modulus of the
correction to the phase velocity increases. Of course, we cannot trust
this calculation for such early times since the correction would
become too large, and since General Relativity would not be valid as
an effective low energy/large distance theory in that regime.

It is worth mentioning that similar results are obtained due to QED
vacuum polarisation~\cite{dh}.  The QED effects are generically much
larger than the graviton corrections. However, there are situations in
which the QED correction vanishes, while the graviton correction does
not.  To show an explicit example, let us consider de Sitter
spacetime.  Virtual $e^-e^+$ pairs modify Maxwell's equations as
follows~\cite{dh}

\begin{equation}
\left( 1+ \frac{7 \alpha R}{1080 \pi m^2} \right) D_{\mu} F^{\mu\nu}=0
\; ,
\end{equation}
where $m$ is the mass of the electron and $\alpha$ the fine structure
constant. The corrected equations coincide with Maxwell's equations up
to a trivial normalization and the dispersion relation is the
classical one.  However, graviton vacuum corrections in de Sitter
spacetime do affect the propagation of individual light rays.


\subsection{STATIC AND SPHERICALLY SYMMETRIC BACKGROUNDS}
\label{schwarzschild}

In this section we consider a static and spherically symmetric
spacetime described by the metric

\begin{equation}
{\rm d}s^2 = -A(r) {\rm d}t^2 + B(r){\rm d}r^2 + r^2 {\rm d}\Omega^2
\; .
\label{ssmetric}
\end{equation}
In order to compute the tensor $M^{\mu\nu\sigma\tau}$ one needs to
calculate the graviton two-point function, evaluated in an arbitrary
quantum state. It would be a rather formidable task to explicitly
calculate such an object. Instead, we will use symmetry arguments and
assume that the graviton quantum state preserves the symmetries of the
background metric to derive the basic structure of the tensor
$M^{\mu\nu\sigma\tau}$. It is shown in Appendix \ref{black-appendix}
that the tensor $M^{\mu\nu\sigma\tau}$ can be written as

\begin{eqnarray}
M^{\mu \nu \sigma \tau} (r)&=&
f_1(r) U^{\mu \nu } U^{\sigma \tau }
+
f_2(r) V^{\mu \nu } V^{\sigma \tau }
+
f_3(r) (X^{\mu \nu } X^{\sigma \tau } + Y^{\mu \nu } Y^{\sigma \tau })
+
f_4(r) (W^{\mu \nu } W^{\sigma \tau } + Z^{\mu \nu } Z^{\sigma \tau })
\; ,
\label{m-blackhole}
\end{eqnarray}
where the functions $f_i(r)$ depend on the particular choice of
vacuum. The antisymmetric order two tensors $U^{\mu \nu }, V^{\mu \nu
}, W^{\mu \nu }, X^{\mu \nu }, Y^{\mu \nu }$, and $Z^{\mu \nu }$ are
defined in Appendix \ref{black-appendix}.  Just as in the case of RW
backgrounds, the only non-vanishing components are $M^{0i0j}$ and
$M^{ijkl}$, with $i,j,k,l$ spatial indices.  In general the structure
of the tensor $M^{\mu\nu\sigma\rho}$ in Eq. (\ref{m-blackhole}) is
much more complicated than that for the Riemann tensor corresponding
to the metric (\ref{ssmetric}).  However, this form for the tensor
$M^{\mu\nu\sigma\tau}$ is good enough to carry an analysis parallel to
that of Drummond and Hathrell~\cite{dh}.

Starting from Eq. (\ref{dhnew}), which describes the propagation of an
EM wave in the presence of gravitons, we show in Appendix
\ref{black-appendix} that it has non-trivial solutions only when the
wave momentum $k_{\mu}$ satisfies the following determinantal
condition

\begin{eqnarray}
&&
k^2
\left[ (1+8f_3)k^2- 8 (f_3+f_4)k_r^2- 8 (f_2+f_3)(k_\theta^2 + k_\phi^2)
\right]
\nonumber
\\
&&\times
\left[ (1+8f_3)(1+8f_1)k^2- 8 (f_3+f_4)(1+8f_1)k_r^2
- 8 (f_1+f_4)(1+8f_3)(k_\theta^2 + k_\phi^2)
\right]
=0
\; .
\label{dispersion}
\end{eqnarray}
The solution $k^2=0$ corresponds to the usual dispersion relation, in
which the light ray follows the null geodesics of the background
metric.  Apart from this (trivial) case, the previous equation admits
new dispersion relations, depending on the direction of propagation
and polarisation of the EM radiation field.

When the light ray moves radially ($k_\theta=k_\phi=0$) the
determinantal condition, (see Eq. (\ref{dispersion})), has two
possible solutions

\begin{mathletters}
\begin{eqnarray}
(1+8f_3)(-k_t^2+k_r^2)- 8 (f_3+f_4)k_r^2&=&0
\; ,
\\
(1+8f_3)(1+8f_1)(-k_t^2 +k_r^2)- 8 (f_3+f_4)(1+8f_1)k_r^2
&=&0
\; .
\end{eqnarray}
\end{mathletters}
If we assume that $(1+8f_1)\neq 0$ the two dispersion relations that
follow from the above equations are the same, which agrees with the
fact that as the gravitational background is spherically symmetric, a
radial EM wave should not be affected by birefringence.  We can
conclude that for radial light rays the absolute value of the quantum
corrected velocity is given by

\begin{eqnarray}
\Big\vert \frac{k_t}{k_r} \Big\vert= 1-4(f_4+f_3)
\; .
\end{eqnarray}

When the EM wave moves transversally ($k_r=k_\theta=0$), we get the
following two possible solutions (for $1+8f_3 \neq 0$)

\begin{mathletters}
\begin{eqnarray}
-(1+8f_3)k_t^2+ (1-8 f_2)k_\phi^2 &=&0
\; ,
 \\
-(1+8f_1)k_t^2 +(1- 8 f_4) k_\phi^2 &=&0
\; .
\end{eqnarray}
\end{mathletters}
As opposed to the previous case, the two dispersion relations that
follow from these two equations are different. Light rays propagate
with different velocities

\begin{mathletters}
\begin{eqnarray}
\Big\vert \frac{k_t}{k_\phi} \Big\vert = 1-4(f_2+f_3)
\; ,
 \\
\Big\vert \frac{k_t}{k_\phi} \Big\vert = 1-4(f_1+f_4)
\; ,
\end{eqnarray}
\end{mathletters}
depending on their polarisation. Similarly, given the symmetry under
the exchange of $\phi$ to $\theta$, azimuthal moving EM waves
($k_r=k_\phi=0$) have the same dispersion relations as transverse
light rays.  Both for traverse and azimuthal propagation we obtain
gravitational birefringence due to graviton vacuum
fluctuations. Similar results are obtained from corrections due to QED
vacuum effects.

It is worth mentioning that the previous results are applicable to any
static and spherically symmetric background. We will now consider two
particular examples, the Schwarzschild background and the
Reissner-Nordstr\"{o}m background~\footnote{As we already mentioned at
the end of Section \ref{equations}, the background metric should be a
solution of the semi-classical Einstein equation. This is not the case
for the particular examples we will be considering in this Section. We
include them only for illustrative purposes.}. The Schwarzschild
spacetime is described by the metric

\begin{equation}
{\rm d}s^2= - \left( 1 - \frac{2 M G}{r} \right) {\rm d}t^2 +
        \left( 1 - \frac{2 M G}{r} \right)^{-1} {\rm d}r^2  +
        r^2 ({\rm d}\theta^2 + \sin^2\theta {\rm d}\phi^2)
\; .
\end{equation}
In this case, using dimensional analysis, we can write the functions
$f_i(r)$ ($i=1,\ldots,4$) as $f_i(r)=(M_{\rm P}/M)^2 {\cal F}_i(r/2M
G)$ with ${\cal F}_i$ four dimensionless functions and $M_{\rm P}$ the
Planck mass. As we have already mentioned, the exact form of these
functions is unknown, since it is not possible to calculate explicitly
the graviton two-point function for the Schwarzschild background.
However, near the horizon $r \approx 2 M G$, we can approximate

\begin{eqnarray}
M^{\mu \nu \sigma \tau} (r \approx2 M G)
& \approx &
\left( \frac{M_{P}}{M}\right)^2
\left[
{\cal F}_1(1) U^{\mu \nu } U^{\sigma \tau }
+
{\cal F}_2(1) V^{\mu \nu } V^{\sigma \tau }
+
{\cal F}_3(1) (X^{\mu \nu } X^{\sigma \tau } + Y^{\mu \nu } Y^{\sigma \tau
})
\right.
\nonumber
\\
&+&
\left.
{\cal F}_4(1) (W^{\mu \nu } W^{\sigma \tau } + Z^{\mu \nu } Z^{\sigma \tau
})
\right]
\; .
\label{m-horizon}
\end{eqnarray}
Hence, near the horizon the birefringence effects induced by gravitons
are of order $O(M_{P}^2 / M^2)$.

The Reissner-Nordstr\"{o}m spacetime describes the metric of a static
and charged black hole

\begin{equation}
{\rm d}s^2= - \left( 1 - \frac{2 M G}{r} + \frac{Q^2 G}{4 \pi r^2} \right)
{\rm d}t^2 +
\left( 1 - \frac{2 M G}{r} + \frac{Q^2 G}{4 \pi r^2} \right)^{-1} {\rm
d}r^2  +
        r^2 ({\rm d}\theta^2 + \sin^2\theta {\rm d}\phi^2)
\; .
\end{equation}
This spacetime has in general two event horizons $r_\pm$. The exterior
one at $r_+=M G [1+ (1-Q^2/4 \pi M^2)^{1/2}]$ coincides with that of
the Schwarzschild metric in the limit $Q^2 \ll M^2$.  We can carry out
an analysis similar to the previous one, and write $f_i(r)=(M_{P}/M)^2
{\cal G}_i(r/2M G, r/ Q \sqrt{G})$ with ${\cal G}_i$ four new
dimensionless functions.  Near the exterior event horizon ($r=r_+$) we
can approximate

\begin{eqnarray}
M^{\mu \nu \sigma \tau} (r \approx r_+)
& \approx &
\left( \frac{M_{P}}{M}\right)^2
\left[
{\cal G}_1 \left( \frac{r_+}{2 M G}, \frac{r_+}{Q \sqrt{G}}
\right)
U^{\mu \nu } U^{\sigma \tau }
+
{\cal G}_2 \left( \frac{r_+}{2 M G}, \frac{r_+}{Q \sqrt{G}}
\right)
V^{\mu \nu } V^{\sigma \tau }
\right. \nonumber \\
&+& \left.
{\cal G}_3 \left(\frac{r_+}{2 M G}, \frac{r_+}{Q \sqrt{G}}
\right)
(X^{\mu \nu } X^{\sigma \tau } + Y^{\mu \nu } Y^{\sigma \tau }) +
{\cal G}_4 \left( \frac{r_+}{2 M G}, \frac{r_+}{Q \sqrt{G} }
\right)
(W^{\mu \nu } W^{\sigma \tau } + Z^{\mu \nu } Z^{\sigma \tau })
\right]
\; .
\end{eqnarray}
Hence, near the outer event horizon the birefringence effects induced
by one-loop gravitons are again of order $O(M_{P}^2 / M^2)$.


\section{CONCLUSIONS}

In this paper we have computed the quantum corrections to (classical)
Maxwell's equations due to the interaction between the EM field and
the fluctuations of the spacetime metric. The modified equations look
like Maxwell's equations in the presence of a linear medium, with
electric permitivity and magnetic permeability tensors proportional to
the coincidence limit of the graviton two-point function. From the
modified equations we have found the quantum corrections to the
dispersion relations. In general, as for the case of linear media in
classical electrodynamics, the speed of light depends on the direction
of propagation and on the state of polarisation.  The quantum
corrections we have computed should be considered along with the
graviton back reaction on the background metric (this is crucial, for
example, for the cancellation of the gauge-breaking dependence; see
Refs.~\cite{dddm,dd}).  Throughout the paper we have assumed that the
metric does contain such (back reaction) corrections, and have focused
on the interaction of the gravitons with the electromagnetic field.

We have shown that, when the EM field wavelength $\lambda$ is small
compared to the typical scale $L_c$ of variation of the permitivity
and permeability tensors, the corrections to the speed of light are
proportional to $(L_P/L_c)^2$.  We have described in detail the
quantum corrections in both RW gravitational backgrounds and static
spherically symmetric spacetimes.

For RW spacetimes we have computed the quantum corrections by two
different methods: the analysis of the modified Maxwell's equations in
a coordinate basis and the study of the dispersion relations in a
local Lorentz frame.  The corrections we found are similar to those of
Ref.~\cite{dd}, where the analysis was based on the effect of
gravitons on the spacetime null geodesics. The results agree
qualitatively but not quantitatively (as is to be expected), since the
coupling of gravitons to point particles is in general different to
the coupling to massless fields.

For spherically symmetric spacetimes we have been able to estimate the
quantum corrections using symmetry and dimensional arguments (see
Appendix \ref{black-appendix}), avoiding the explicit computation of
the graviton two-point function.

The quantum corrections for small wavelengths are also qualitatively
similar to those produced by virtual $e^-e^+$ in non-trivial
backgrounds.  However, for some particular cases (as de Sitter
background), the QED vacuum does not affect the propagation of
photons, whereas the graviton vacuum does induce a modification on the
propagation of light rays.

The opposite limit, $\lambda \gg L_c$, is more interesting from a
phenomenological point of view.  Assuming a non-trivial and random
spacetime structure at scales of order $L_c$, the modified field
equations are similar to the ones describing the propagation of
classical waves in random media. To lowest order, it is possible to
describe the wave propagation in terms of an ``effective medium''. The
average corrections to the speed of light are independent of the
wavelength and proportional to $(L_P/L_c)^2$.  The ``effective
medium'' is only an averaged description. Stochastic fluctuations are
expected to occur, for example, in the arrival time of photons coming
from point sources~\cite{ellismavronanop,ford1}.  To next order, the
corrections are proportional to $L_P^2 E/(L_c\lambda)$, where $E$ is
the photon energy . In the extreme case $L_P\sim L_c$ (which we cannot
reach within our effective field theory approach) the corrections
would be proportional to $E/E_P$, where $E_P$ is the Planck energy.
In this regime quantum gravity effects increase with energy, while
other medium effects should decrease with energy. They can be
distinguished by this property and could be relevant in cosmological
situations~\cite{nature}.

In the effective field theory approach to quantum gravity one also
expects classical, stochastic fluctuations of the spacetime
metric. Its dynamics should be described by a ``semi-classical
Einstein-Langevin equation"~\cite{el}.  These classical fluctuations
will also affect the propagation of photons (see for
example~\cite{bltom}).


\section{Acknowledegments}

{We thank the Universidad Carlos III de Madrid, and particularly
Dr. Grant D. Lythe for his hospitality during D.D. visit to Spain,
while part of this work was carried out.  We also thank Paul
R. Anderson for helpful discussions and a careful reading of the
manuscript.  F.D.M.  acknowledges the support from Universidad de
Buenos Aires and CONICET (Argentina) and C.M.-P. from CAB (CSIC/INTA)
(Spain).}


\appendix

\section{Structure of the tensor M for homogeneous and isotropic
backgrounds}
\label{rw-appendix}

We consider a flat homogeneous and isotropic spacetime described by
the metric (conformal coordinates)

\begin{equation}
{\rm d}s^2 = a^2(\eta) (-{\rm d}\eta^2 + {\rm d}x^2 + {\rm d}y^2
+ {\rm d}z^2)
\; .
\end{equation}
In order to compute the tensor $M^{\mu\nu\sigma\tau}$ one needs to
calculate the graviton two-point function, evaluated in an arbitrary
quantum state.  We will assume that the graviton state preserves the
symmetries of the background metric. The aim of this appendix is to
use those symmetries to find the structure of the tensor $M^{\mu \nu
\sigma \tau}$ given in Eq. (\ref{eme}). In order to obtain this
structure we choose an orthonormal basis of the spacetime under
consideration. The orthonormal basis is defined by the vector fields

\begin{mathletters}
\begin{eqnarray}
{\bf e}_{\eta}
&=&
\frac{1}{a(\eta)} \frac{\partial}{\partial \eta}
\; ,
\\
{\bf e}_x
&=&
\frac{1}{a(\eta)}
 \frac{\partial}{\partial x}
\; ,
\\
{\bf e}_y
&=&
\frac{1}{a(\eta)}
 \frac{\partial}{\partial y}
\; ,
\\
{\bf e}_z
&=&
\frac{1}{a(\eta)}
 \frac{\partial}{\partial z}
\; .
\label{baseRW}
\end{eqnarray}
\end{mathletters}
We also introduce the following set of antisymmetric tensors

\begin{mathletters}
\begin{eqnarray}
U_x^{\mu \nu }
&=& {\bf e}_{\eta}^\mu {\bf e}_x^\nu -{\bf e}_{\eta}^\nu {\bf e}_x^\mu
\; ,
\\
U_y^{\mu \nu }
&=& {\bf e}_{\eta}^\mu {\bf e}_y^\nu -{\bf e}_{\eta}^\nu {\bf e}_y^\mu
\; ,
\\
U_z^{\mu \nu }
&=& {\bf e}_{\eta}^\mu {\bf e}_z^\nu -{\bf e}_{\eta}^\nu {\bf e}_z^\mu
\; ,
\\
V_x^{\mu \nu }
&=& {\bf e}_y^\mu {\bf e}_z^\nu -{\bf e}_y^\nu {\bf e}_z^\mu
\; ,
\\
V_y^{\mu \nu }
&=& {\bf e}_z^\mu {\bf e}_x^\nu -{\bf e}_z^\nu {\bf e}_x^\mu
\; ,
\\
V_z^{\mu \nu }
&=& {\bf e}_x^\mu {\bf e}_y^\nu -{\bf e}_x^\nu {\bf e}_y^\mu
\; .
\end{eqnarray}
\end{mathletters}
Given the fact that the ${\bf e}$'s form an orthonormal basis, this
set of six tensors constitutes a basis for the antisymmetric order two
tensors.

It is easy to see that the symmetries of the tensor $M^{\mu \nu \sigma
\tau}$ (antisymmetric in the first two indices, and the last two, and
symmetric under the exchange of the two pair of indices) imply that
the tensor must be a linear combination of the following kind (with
coefficients that may depend only on $\eta$)

\begin{eqnarray}
M^{\mu \nu \sigma \tau}&=&
a_1(\eta) U_x^{\mu \nu }   U_x^{\sigma \tau }
+ a_2(\eta)
(U_x^{\mu \nu }   U_y^{\sigma \tau } +U_y^{\mu \nu }   U_x^{\sigma \tau } )
+ a_3(\eta)
( U_x^{\mu \nu }   U_z^{\sigma \tau } + U_z^{\mu \nu }   U_x^{\sigma \tau }
)
\nonumber
\\
&+&
a_4(\eta)
(U_x^{\mu \nu }   V_x^{\sigma \tau } +V_x^{\mu \nu }   U_x^{\sigma \tau } )
+ a_5(\eta)
( U_x^{\mu \nu }   V_y^{\sigma \tau } +
 V_y^{\mu \nu }   U_x^{\sigma \tau } )
+ a_6(\eta) (U_x^{\mu \nu }   V_z^{\sigma \tau }
+V_z^{\mu \nu }   U_x^{\sigma \tau } )
\nonumber
\\
&+&
a_7(\eta) U_y^{\mu \nu }   U_y^{\sigma \tau }
+ a_8(\eta)( U_y^{\mu \nu }   U_z^{\sigma \tau }
+ U_z^{\mu \nu }   U_y^{\sigma \tau } )
+ a_9(\eta)( U_y^{\mu \nu }   V_x^{\sigma \tau }
+ V_x^{\mu \nu }   U_y^{\sigma \tau } )
+ a_{10}(\eta)
( U_y^{\mu \nu }   V_y^{\sigma \tau }
+ V_y^{\mu \nu }   U_y^{\sigma \tau } )
\nonumber
\\
&+&
 a_{11}(\eta)( U_y^{\mu \nu }   V_z^{\sigma \tau }
+ V_z^{\mu \nu }   U_y^{\sigma \tau } )
+ a_{12}(\eta) U_z^{\mu \nu }   U_z^{\sigma \tau }
+ a_{13}(\eta)( U_z^{\mu \nu }   V_x^{\sigma \tau }
+
V_x^{\mu \nu }   U_z^{\sigma \tau } )
\nonumber
\\
&+&
 a_{14}(\eta)( U_z^{\mu \nu }   V_y^{\sigma \tau }
+ V_y^{\mu \nu }   U_z^{\sigma \tau } )
+ a_{15}(\eta) (U_z^{\mu \nu }   V_z^{\sigma \tau }
+V_z^{\mu \nu }   U_z^{\sigma \tau } )
+ a_{16}(\eta) V_x^{\mu \nu }   V_x^{\sigma \tau }
\nonumber
\\
&+&
 a_{17}(\eta)( V_x^{\mu \nu }   V_y^{\sigma \tau }
+ V_y^{\mu \nu }   V_x^{\sigma \tau } )
+ a_{18}(\eta)
( V_x^{\mu \nu }   V_z^{\sigma \tau } + V_z^{\mu \nu }   V_x^{\sigma \tau
})
\nonumber
\\
&+&
 a_{19}(\eta) V_y^{\mu \nu }   V_y^{\sigma \tau }
+ a_{20}(\eta)( V_y^{\mu \nu }   V_z^{\sigma \tau }
+ V_z^{\mu \nu }   V_y^{\sigma \tau } )
+ a_{21}(\eta) V_z^{\mu \nu }   V_z^{\sigma \tau }  \,.
\label{longeme-rw}
\end{eqnarray}
Since the metric is homogeneous and isotropic the functions
$a_2,a_3,a_4,a_5,a_6,a_8,a_9,a_{10},a_{11},
a_{13},a_{14},a_{15},a_{17},a_{18}$, and $a_{20}$ must vanish
identically.  Here we should stress once again the fact that the
quantum state of the gravitons does not break these symmetries. The
background symmetries also imply that $a_1=a_7=a_{12} \equiv \alpha$
and $a_{16}=a_{19}=a_{21} \equiv \beta$.

After these symmetry considerations, we can finally write the general
expression of the tensor $M^{\mu\nu\sigma\tau}$ for a homogeneous and
isotropic background

\begin{eqnarray}
M^{\mu \nu \sigma \tau}&=& \alpha(\eta)
( U_x^{\mu \nu }   U_x^{\sigma \tau }
+
U_y^{\mu \nu }   U_y^{\sigma \tau } +
U_z^{\mu \nu }   U_z^{\sigma \tau } )
+ \beta(\eta)
( V_x^{\mu \nu }   V_x^{\sigma \tau }
+
V_y^{\mu \nu }   V_y^{\sigma \tau } +
V_z^{\mu \nu }   V_z^{\sigma \tau } )
\; .
\label{emeapp-rw}
\end{eqnarray}
This is all the information we can obtain regarding the tensor
structure of $M^{\mu\nu\sigma\tau}$ by making use of the symmetries of
the gravitational background. The functions $\alpha(\eta)$ and
$\beta(\eta)$ will depend on the particular choice of vacuum and are
not known a priori.

We now make use of the newly obtained form for $M^{\mu\nu\sigma\tau}$,
(see equation (\ref{emeapp-rw})) to solve the equation of motion for
the EM field in the presence of one-loop quantum fluctuations in flat
homogeneous and isotropic backgrounds (see Eq. (\ref{dhnew})).
Following the very same steps as described in Ref.~\cite{dh}, we
introduce the six (dependent) functions

\begin{eqnarray}
u_x = f_{\mu \nu}U_x^{\mu \nu}   & ~~;~~ &
v_x = f_{\mu \nu}V_x^{\mu\nu}
;
\nonumber \\
u_y = f_{\mu \nu}U_y^{\mu\nu}
& ~~;~~ &
v_y = f_{\mu \nu}V_y^{\mu \nu}
;
\nonumber \\
u_z = f_{\mu \nu}U_z^{\mu \nu}
 & ~~;~~ &
v_z = f_{\mu \nu}V_z^{\mu \nu}.
\end{eqnarray}
{From} Eq. (\ref{max2}) it follows that

\begin{equation}
f_{\mu \nu} = k_\mu a_\nu - k_\nu a_\mu
\; ,
\end{equation}
for some gauge potential $a_{\mu}$. Hence, for a given EM wave
momentum $k_{\mu}$, $f_{\mu\nu}$ has three independent components (one
amplitude and two polarisations), as we still have the choice of gauge
for the EM field. Since $f_{\mu\nu}$ is gauge invariant, without loss
of generality we consider the Coulomb gauge and choose
$a_{\eta}=0$. With this choice the three non-vanishing components for
the gauge potential are $a_x$, $a_y$, and $a_z$, so that

\begin{eqnarray}
f_{{\eta} x} = k_{\eta} a_x
& ~~;~~ &
f_{xy} = k_x a_y - k_y a_x ;
 \nonumber \\
f_{{\eta} y} = k_{\eta} a_y
& ~~;~~ &
f_{yz} = k_y a_z - k_z a_y
;
\nonumber \\
f_{{\eta} z} = k_{\eta} a_z
& ~~;~~ &
f_{zx} = k_z a_x - k_x a_z ,
\end{eqnarray}
and we can then write

\begin{eqnarray}
u_x = 2 f_{{\eta} x}
&~~;~~&
v_x = 2 f_{yz}
;
\nonumber \\
u_y =2 f_{{\eta} y}
&~~;~~&
v_y = 2 f_{zx}
;
\nonumber \\
u_z = 2f_{{\eta} z}
&~~;~~&
v_z=2 f_{xy} ,
\end{eqnarray}
so that in terms of the independent set $\left\{u_x,u_y,u_z \right\}$
the three dependent ones can be written as

\begin{mathletters}
\begin{eqnarray}
v_x&=& \frac{1}{k_{\eta}} (k_y u_z - k_z u_y)
\; ,
\\
v_y&=& \frac{1}{k_{\eta}} (k_z u_x - k_x u_z)
\; ,
\\
v_z&=& \frac{1}{k_{\eta}} (k_x u_y - k_y u_x)
\; .
\end{eqnarray}
\end{mathletters}
With all these definitions in mind we project equation (\ref{dhnew})
onto the three tensors $U_x$, $U_y$, and $U_z$ (that yield $u_x,u_y,$
and $u_z$, respectively), to obtain the following set of equations

\begin{mathletters}
\begin{eqnarray}
0 &=&
k^2 u_x
+ 8 \alpha
[u_x(-k_{\eta}^2+k_x^2) + k_x(k_y u_y + k_z u_z)]
+ 8 \beta k_{\eta} (k_y v_z -k_z v_y)
\; ,
\\
0 &=&
k^2 u_y
+ 8 \alpha
[u_y(-k_{\eta}^2+k_y^2) + k_y(k_x u_x + k_z u_z)]
+ 8 \beta k_{\eta} (k_z v_x -k_x v_z)
\; ,
\\
0 &=&
k^2 u_z
+ 8 \alpha
[u_z(-k_{\eta}^2+k_z^2) + k_z(k_x u_x + k_y u_y)]
+ 8 \beta k_{\eta} (k_x v_y -k_y v_x)
\; .
\end{eqnarray}
\end{mathletters}
We point out that the components of the vector $k$ correspond to the
orthonormal basis given in Eqs. (A2), so that $k=k^\eta {\bf e}_\eta +
k^x {\bf e}_x + k^y {\bf e}_y +k^z {\bf e}_z $.  If we write the
components ($v_x,v_y,v_z$) as functions of ($u_x,u_y,u_z$), we have

\begin{mathletters}
\begin{eqnarray}
0 &=&
k^2 u_x
+ 8 \alpha
[u_x(-k_{\eta}^2+k_x^2) + k_x(k_y u_y + k_z u_z)]
+ 8 \beta[ k_y( k_x u_y -k_y u_x)-k_z( k_z u_x -k_x u_z)
]
\; ,
\\
0 &=&
k^2 u_y
+ 8 \alpha
[u_y(-k_{\eta}^2+k_y^2) + k_y(k_x u_x + k_z u_z)]
+ 8 \beta[ k_z( k_y u_z -k_z u_y)-k_x( k_x u_y -k_y u_x)
]
\; ,
\\
0 &=&
k^2 u_z
+ 8 \alpha
[u_z(-k_{\eta}^2+k_z^2) + k_z(k_x u_x + k_y u_y)]
+ 8 \beta[ k_x( k_z u_x -k_x u_z)-k_y( k_y u_z -k_z u_y)
]
\; .
\end{eqnarray}
\end{mathletters}
Imposing the condition that the determinant of this set of equations
vanishes, so that we do not obtain the trivial solution
($u_x=u_y=u_z=0$), we obtain the following determinantal restriction

\begin{eqnarray}
k^2 (1+8\alpha)
\left[ (1+8\alpha)k^2- 8 (\alpha+\beta)(k_x^2+k_y^2 + k_z^2)
\right]^2
&=&0
\;  .
\end{eqnarray}

Let us assume that the EM radiation is characterized by a
three-dimensional momentum ${\bf k}$, such that $k_x^2+k_y^2 + k_z^2=
{\bf k} \cdot {\bf k}$. The non-trivial dispersion relation becomes
then

\begin{eqnarray}
 (1+8\alpha)(-k_{\eta}^2+{\bf k} \cdot {\bf k})
- 8 (\alpha+\beta){\bf k} \cdot {\bf k}
&=&0
\;  .
\end{eqnarray}
The EM radiation waves will propagate with the following dispersion
relation

\begin{eqnarray}
 \frac{k_{\eta}^2}{{\bf k} \cdot {\bf k}}  = 1-8(\alpha+\beta)
\; ,
\end{eqnarray}
or equivalently
\begin{eqnarray}
\Big\vert
 \frac{k_{\eta}}{{\bf k}}
\Big\vert = 1-4(\alpha+\beta)
\; .
\end{eqnarray}
We can now compare with the previous formulation of this problem in
terms of the two-point function of the gravitons (see section
\ref{RW}). We already know from symmetry considerations that the
tensor $M^{\mu\nu\sigma\tau}$ can be written as

\begin{eqnarray}
M^{\mu \nu \sigma \tau}&=& \alpha({\eta})
( U_x^{\mu \nu }   U_x^{\sigma \tau }
+
U_y^{\mu \nu }   U_y^{\sigma \tau } +
U_z^{\mu \nu }   U_z^{\sigma \tau } )
+
 \beta({\eta})
( V_x^{\mu \nu }   V_x^{\sigma \tau }
+
V_y^{\mu \nu }   V_y^{\sigma \tau } +
V_z^{\mu \nu }   V_z^{\sigma \tau } )
\; .
\end{eqnarray}

Let us now calculate the only non-vanishing components of this tensor
($M^{\eta i \eta j}$ and $M^{ijmn}$) in the coordinate basis defined
by the vector fields $\partial_{\eta}, \partial_x, \partial_y$ and
$\partial_z$.  We get

\begin{eqnarray}
M^{\eta i \eta j}
&=& \alpha(\eta)
( U_x^{\eta i}   U_x^{\eta j}
+
U_y^{\eta i}   U_y^{\eta j } +
U_z^{\eta i }   U_z^{\eta j } )
=\frac{\alpha(\eta)}{a^2(\eta)} g^{ij} ,
\\
M^{ijmn}&=&
 \beta(\eta)
( V_x^{ij }   V_x^{mn }
+
V_y^{ij }   V_y^{mn } +
V_z^{ij }   V_z^{mn } )
= \beta(\eta) (g^{im} g^{jn} - g^{in} g^{jm})
\; ,
\end{eqnarray}
which means in particular that the functions $\alpha(\eta)$ and
$\beta(\eta)$ introduced in this appendix are related to the functions
defined in Eqs. (\ref{eme1}) and (\ref{eme2}) as
$f_1(\eta)=\alpha(\eta)/a^2(\eta)$ and $f_2(\eta)=\beta(\eta)$.
Hence, we conclude that

\begin{eqnarray}
\Big\vert
 \frac{k_{\eta}}{{\bf k}}
\Big\vert =  1 - 4 (\alpha+\beta) = 1 - 4 [ a^2(\eta) f_1(\eta) + f_2(\eta)
]
= 1 + \frac{\kappa^2}{6} \langle \phi^2(\eta) \rangle
\; ,
\end{eqnarray}
which agrees with our previous result obtained from Maxwell's
equations and the two-point function of the gravitons (see equation
(\ref{veff})).


\section{Structure of the tensor M for static and spherically symmetric
backgrounds}
\label{black-appendix}

We assume a static and spherically symmetric spacetime described by
the metric

\begin{equation}
{\rm d}s^2 = -A(r) {\rm d}t^2 + B(r){\rm d}r^2 + r^2 {\rm d}\Omega^2
\; .
\end{equation}
In order to compute the general form of the tensor
$M^{\mu\nu\sigma\tau}$ we will follow the same steps as in Appendix
\ref{rw-appendix}.  We choose an orthonormal basis of the spacetime
given by the vectors ${\bf e}_t$, ${\bf e}_r$, ${\bf e}_\theta$, and
${\bf e}_\phi$.  This orthonormal basis is defined by

\begin{mathletters}
\begin{eqnarray}
{\bf e}_t
&=&
\left[A(r)\right]^{-1/2}
\frac{\partial}{\partial t}
\; ,
\\
{\bf e}_r
&=&
\left[B(r)\right]^{-1/2}
 \frac{\partial}{\partial r}
\; ,
\\
{\bf e}_\theta
&=&
\frac{1}{r} \frac{\partial}{\partial \theta}
\; ,
\\
{\bf e}_\phi
&=&
\frac{1}{r \sin \theta} \frac{\partial}{\partial \phi}
\; .
\end{eqnarray}
\end{mathletters}
We also introduce the following basis of antisymmetric order two
tensors

\begin{mathletters}
\begin{eqnarray}
U^{\mu \nu }
&=& {\bf e}_t^\mu {\bf e}_r^\nu -{\bf e}_t^\nu {\bf e}_r^\mu
\; ,
\\
V^{\mu \nu } &=&
{\bf e}_\theta^\mu {\bf e}_\phi^\nu -{\bf e}_\theta^\nu {\bf e}_\phi^\mu
\; ,
\\
X^{\mu \nu } &=&
{\bf e}_t^\mu {\bf e}_\theta^\nu -{\bf e}_t^\nu {\bf e}_\theta^\mu
\; ,
\\
Y^{\mu \nu } &=&
{\bf e}_t^\mu {\bf e}_\phi^\nu -{\bf e}_t^\nu {\bf e}_\phi^\mu
\; ,
\\
W^{\mu \nu } &=&
{\bf e}_r^\mu {\bf e}_\theta^\nu -{\bf e}_r^\nu {\bf e}_\theta^\mu
\; ,
\\
Z^{\mu \nu } &=&
{\bf e}_r^\mu {\bf e}_\phi^\nu -{\bf e}_r^\nu {\bf e}_\phi^\mu
\; .
\end{eqnarray}
\end{mathletters}
The tensor $M^{\mu \nu \sigma \tau}$ must be a linear combination of
the following kind (with coefficients that may depend only on $r$)

\begin{eqnarray}
M^{\mu \nu \sigma \tau}&=&
a_1(r)
 U^{\mu \nu }   U^{\sigma \tau } +
a_2(r) (U^{\mu \nu }   V^{\sigma \tau} +  V^{\mu \nu }   U^{\sigma \tau })
+ a_3(r) (U^{\mu \nu }   X^{\sigma \tau} + X^{\mu \nu }   U^{\sigma \tau})
\nonumber
\\
&+&a_4(r) ( U^{\mu \nu }   Y^{\sigma \tau } + Y^{\mu \nu }   U^{\sigma \tau
})
+ a_5(r) ( U^{\mu \nu }   W^{\sigma \tau } + W^{\mu \nu }   U^{\sigma \tau
})
+ a_6(r) ( U^{\mu \nu }   Z^{\sigma \tau } + Z^{\mu \nu }   U^{\sigma \tau
})
\nonumber
\\
&+& a_7(r)V^{\mu \nu }   V^{\sigma \tau }
+ a_8(r) (V^{\mu \nu }   X^{\sigma \tau } + X^{\mu \nu }  V^{\sigma \tau }
)
+ a_9(r) (V^{\mu \nu }   Y^{\sigma \tau } + Y^{\mu \nu }   V^{\sigma \tau
})
\nonumber
\\
&+&
a_{10}(r) (V^{\mu \nu }   W^{\sigma \tau } + W^{\mu \nu }   V^{\sigma \tau
})
+a_{11}(r) (V^{\mu \nu }   Z^{\sigma \tau } + Z^{\mu \nu }  V^{\sigma \tau
})
+a_{12}(r) X^{\mu \nu }   X^{\sigma \tau }
\nonumber
\\
&+& a_{13}(r) (X^{\mu \nu } Y^{\sigma \tau } + Y^{\mu \nu } X^{\sigma \tau
})
+a_{14}(r) (X^{\mu \nu }   W^{\sigma \tau } + W^{\mu \nu }  X^{\sigma \tau
})
+a_{15}(r) (X^{\mu \nu }   Z^{\sigma \tau } + Z^{\mu \nu }  X^{\sigma \tau
})
\nonumber
\\
&+&a_{16}(r) Y^{\mu \nu }   Y^{\sigma \tau }
+a_{17}(r) ( Y^{\mu \nu }   W^{\sigma \tau } + W^{\mu \nu } Y^{\sigma \tau
} )
+ a_{18}(r) (Y^{\mu \nu }   Z^{\sigma \tau } + Z^{\mu \nu } Y^{\sigma \tau
} )
\nonumber
\\
&+&a_{19}(r) W^{\mu \nu }   W^{\sigma \tau }
+a_{20}(r) (W^{\mu \nu }   Z^{\sigma \tau } + Z^{\mu \nu }  W^{\sigma \tau
} )
+a_{21}(r) Z^{\mu \nu }   Z^{\sigma \tau }\, .
\label{longeme}
\end{eqnarray}
Since the metric is static, we can make use of the time inversion
invariance to show that the terms in Eq. (\ref{longeme}) involving the
functions $a_2,a_5,a_6,a_8,a_9,a_{14},a_{15},a_{17}$, and $a_{18}$
must vanish identically. Here we should stress once again that the
quantum state of the gravitons does not break time inversion
invariance. (This is not true in general. For example, the Unruh
vacuum state in Schwarzschild spacetime does break the time inversion
symmetry.)

Because of the spherical symmetry of the metric (spatial inversion as
well) the following coefficients have to vanish:
$a_3,a_4,a_{10},a_{11},a_{13},$ and $a_{20}$.  Furthermore, the
coefficients $a_{12}$ and $a_{16}$ must be equal as they have to be
invariant under spatial rotations. The same is true for the pair
$a_{19}$ and $a_{21}$.

After these symmetry considerations, we can finally write the general
expression for the tensor $M^{\mu\nu\sigma\tau}$ in a static,
spherically symmetric background

\begin{eqnarray}
M^{\mu \nu \sigma \tau}&=& f_1(r) U^{\mu \nu }   U^{\sigma \tau } +
 f_2(r) V^{\mu \nu }   V^{\sigma \tau }
+
f_3(r) (X^{\mu \nu } X^{\sigma \tau } + Y^{\mu \nu } Y^{\sigma \tau })
+
f_4(r) (W^{\mu \nu } W^{\sigma \tau } + Z^{\mu \nu } Z^{\sigma \tau })
\; .
\label{emeapp}
\end{eqnarray}
The functions $f_i(r)$ (with $i=1,\ldots,4$) will depend on the
particular choice of graviton state and are not known a priori.

As in Apendix A we introduce the six functions

\begin{eqnarray}
u = f_{\mu \nu}U^{\mu \nu} & ~~;~~ & v = f_{\mu \nu}V^{\mu \nu} ; \nonumber
\\
x = f_{\mu \nu}X^{\mu \nu} & ~~;~~ & y = f_{\mu \nu}Y^{\mu \nu} ; \nonumber
\\
w = f_{\mu \nu}W^{\mu \nu} & ~~;~~ & z = f_{\mu \nu}Z^{\mu \nu}.
\end{eqnarray}
In the Coulomb gauge, the three non-vanishing components for the gauge
potential are $a_r$, $a_\theta$, and $a_\phi$, so that

\begin{eqnarray}
f_{t r} = k_t a_r  & ~~;~~ & f_{t \theta} = k_t a_\theta; \nonumber \\
f_{t \phi} = k_t a_\phi & ~~;~~ & f_{r \theta} = k_r a_\theta - k_\theta
a_r ;
\nonumber \\
f_{r \phi} = k_r a_\phi - k_\phi a_r & ~~;~~ &
f_{\theta \phi} = k_\theta a_\phi - k_\phi a_\theta ,
\end{eqnarray}
and we can then write

\begin{eqnarray}
u = 2 f_{t r} &~~;~~& x =2 f_{t \theta}
;
\nonumber \\
y = 2f_{t \phi} &~~;~~& w = 2 f_{r \theta}
;
\nonumber \\
z = 2 f_{r \phi} &~~;~~& v=2 f_{\theta \phi} ,
\end{eqnarray}
so that in terms of the independent set $\left\{u,x,y \right\}$, the
three dependent ones can be written as

\begin{mathletters}
\begin{eqnarray}
w&=& \frac{1}{k_t} (k_r x - k_\theta u)
\; ,
\\
z&=& \frac{1}{k_t} (k_r y - k_\phi u)
\; ,
\\
v&=& \frac{1}{k_t} (k_\theta y - k_\phi x)
\; .
\end{eqnarray}
\end{mathletters}
We now project equation (\ref{dhnew}) onto the three tensors $U$, $X$,
and $Y$ (that yield $u,x,$ and $y$, respectively), to obtain the
following set of equations

\begin{mathletters}
\begin{eqnarray}
0 &=&
k^2 u
- 8 f_1 u (k_t^2 - k_r^2) + 8 f_3 k_r (xk_\theta + y k_\phi)
+ 8 f_4 k_t (w k_\theta + z k_\phi)
\; ,
\\
0 &=&
k^2 x
+ 8 f_1 u k_r k_\theta + 8 f_2 v k_t k_\phi
- 8 f_3 x (k_t^2 -k_\theta^2)
+ 8 f_3 y  k_\theta k_\phi - 8 f_4 w k_t k_r
\; ,
\\
0 &=&
k^2 y
+ 8 f_1 u k_r k_\phi -8 f_2 v k_t k_\theta
+ 8 f_3 x k_\theta k_\phi
- 8 f_3 y  (k_t^2 -k_\phi^2) - 8 f_4 z k_t k_r
\; .
\end{eqnarray}
\end{mathletters}
If we write the components ($v,w,z$) as functions of ($u,x,y$), we
have

\begin{mathletters}
\begin{eqnarray}
0 &=&
\left[ k^2 (1+ 8 f_1)
- 8 (f_1 + f_4) (k_\theta^2 + k_\phi^2)
\right] u
+ 8 (f_3+ f_4) k_r k_\theta  x
+ 8 (f_3+ f_4) k_r k_\phi y
\; ,
\\
0 &=&
 8 (f_1+ f_4) k_r k_\theta  u
+
\left[ k^2 (1+ 8 f_3)
- 8 (f_2 + f_3)  k_\phi^2 - 8 (f_3 + f_4)  k_r^2
\right] x
+ 8 (f_3+f_4)   k_\theta k_\phi y
\; ,
\\
0 &=&
 8 (f_1+ f_4) k_r k_\phi  u
+ 8 (f_2+f_3)   k_\theta k_\phi x
+
\left[ k^2 (1+ 8 f_3)
- 8 (f_2 + f_3)  k_\theta^2 - 8 (f_3 + f_4)  k_r^2
\right] y .
\end{eqnarray}
\end{mathletters}
In this case the components of the vector $k$ in the orthonormal basis
are given by $k=k^t {\bf e}_t + k^r {\bf e}_r + k^\theta {\bf
e}_\theta +k^\phi {\bf e}_\phi $.

Imposing that the determinant of the set of three equations vanishes,
we obtain the following determinantal condition

\begin{eqnarray}
&&
k^2
\left[ (1+8f_3)k^2 - 8 (f_3+f_4)k_r^2 - 8 (f_2+f_3)(k_\theta^2 + k_\phi^2)
\right]
\nonumber
\\
&&\times
\left[ (1+8f_3)(1+8f_1)k^2 - 8 (f_3+f_4)(1+8f_1)k_r^2
- 8 (f_1+f_4)(1+8f_3)(k_\theta^2 + k_\phi^2)
\right]
=0 .
\end{eqnarray}



\begin{references}

\bibitem{adler}
S.L. Adler, Ann. Phys. (N.Y.) {\bf 67}, 599 (1971).

\bibitem{dittrich}
W. Dittrich and H. Gies, Phys. Rev. D {\bf 58},
025004 (1998).

\bibitem{scharn1}
K. Scharnhorst,
Phys. Lett. B {\bf 236}, 354 (1990).

\bibitem{barton}
G. Barton, Phys. Lett. B {\bf 237}, 559 (1990).

\bibitem{dh}
I.T. Drummond and S.J. Hathrell,
Phys. Rev. D {\bf 22}, 343 (1980).

\bibitem{danshore}
R.D. Daniels and G.M. Shore,
Nucl. Phys. B {\bf 425}, 634 (1994).

\bibitem{tarrach}
J.I. Latorre, P. Pascual and R. Tarrach,
Nucl. Phys. B {\bf 437}, 60 (1995).

\bibitem{scharn2}
K. Scharnhorst, talk delivered at the workshop
``Superluminal (?) velocities'', Cologne (1998); hep-th/9810221.

\bibitem{roberto}
D. Bakalov {\it et.al.}, Quantum Semiclass. Opt. {\bf 10}, 239 (1998).

\bibitem{shore}
G.M. Shore, Nucl. Phys. B {\bf 460}, 379 (1996).

\bibitem{donoghue}
J.F. Donoghue, Phys. Rev. D {\bf 50}, 3874 (1994).

\bibitem{newtpot}
J.F. Donoghue, Phys. Rev. Lett. {\bf 72}, 2996 (1994);
I. Muzinich and S. Vokos, Phys. Rev. D {\bf 52}, 3472 (1995);
H. Hamber and
S. Liu, Phys. Lett. B {\bf 357}, 51 (1995).

\bibitem{dddm}
D.A.R. Dalvit and F.D. Mazzitelli, Phys. Rev. D {\bf 56},
7779 (1997).

\bibitem{nature} G. Amelino-Camelia {\it et. al.}, Nature (London) {\bf
393},
763 (1998).

\bibitem{gambinipullin} R. Gambini and J. Pullin, Phys. Rev. D {\bf 59},
124021 (1999).

\bibitem{dewitt}
B.S. Dewitt, ``The Space-Time Approach to Quantum Field Theory'', {\it In
Les
Houches 1983, Proceedings, Relativity, Groups and Topology, 381-738.}

\bibitem{dd}
D.A.R. Dalvit and F.D. Mazzitelli, Phys. Rev. D {\bf 60},
084018 (1999).

\bibitem{sidm}
K. Kazakov and P. Pronin, Nucl. Phys. B {\bf 573}, 536 (2000);
Phys. Rev. D {\bf 62 }, 044043 (2000);
S. Iguri and F.D. Mazzitelli, in preparation.

\bibitem{weave}
A. Ashtekar, C. Rovelli, and L. Smolin, Phys. Rev. Lett.
{\bf 69}, 237 (1992).

\bibitem{allen}
B. Allen, Nucl. Phys. B {\bf 287}, 743 (1987).

\bibitem{fp}
L.H. Ford and L. Parker, Phys. Rev. D {\bf 16}, 1601 (1977).

\bibitem{bunch}
T.S. Bunch, J. Phys. A: Math. Gen. {\bf 13}, 1297 (1980).

\bibitem{BD} N.D. Birrell and P.C.W. Davies, {\it Quantum Fields in Curved
Space} (Cambridge University Press, London, 1982).

\bibitem{ellismavronanop} J. Ellis, N.E. Mavromatos, and D.V.
Nanopoulos, Gen. Rel. and Gravitation {\bf 32}, 127 (2000).

\bibitem{ford1}
L.H. Ford and N.F. Svaiter, Phys.Rev. D {\bf 56},
2226 (1997).

\bibitem{el} F.C. Lombardo and F.D. Mazzitelli, Phys. Rev. D {\bf 55}, 3889
(1997); E. Calzetta, A. Campos, and E. Verdaguer, {\it ibid} {\bf 56}, 2163
(1997).

\bibitem{bltom} B.L. Hu and K. Shiokawa, Phys. Rev. D {\bf 57},
3474 (1998); K. Shiokawa, {\it ibid} {\bf 62}, 024002 (2000).

\end{references}
\end{document}